\documentclass[preprint,secnumarabic,amssymb, nofootinbib, aps, prd,11pt]{revtex4-1}
\usepackage{graphicx}				% Use pdf, png, jpg, or eps§ with pdflatex; use eps in DVI mode
\usepackage{hyperref}
\usepackage{amssymb}
\usepackage{amstext}
\usepackage{setspace}

\newcommand{\be}{\begin{eqnarray}}
\newcommand{\ee}{\end{eqnarray}}

\newcommand{\ma}{\mathcal{A}}
\newcommand{\as}{\mathcal{A}(1,2,\sigma)}
\newcommand{\mb}{\mathcal{B}}

\begin{document}

\title{Dyck Words and Multi-Quark Primitive Amplitudes}%

\author{Tom Melia}%
\email{thomas.melia@cern.ch}
\affiliation{CERN Theory Division, CH-1211,\\
 Geneva 23, Switzerland.}
%\date{August 10, 2010}%
\preprint{CERN-PH-TH/2013-088}%

\begin{abstract}
{I study group theory (Kleiss-Kuijf) relations between purely multi-quark primitive amplitudes at tree level, and 
 prove that they reduce the number of independent primitives 
 to  $(n-2)!/(n/2)!$, where $n$ is the number
of quarks plus antiquarks, in the case where quark lines have different flavours. 
I give an explicit example of an independent basis of
 primitives for any $n$ which is of the form $\ma(1,2,\sigma)$, where $\sigma$ is a permutation based on a Dyck word. }
\end{abstract}

\maketitle
\section{Introduction}

Colour decompositions have proven to be an extremely useful tool
in state-of-the-art calculations of multi-jet cross sections at particle colliders (for a review see e.g.
 \cite{Ellis:2011cr}).
They allow for the definition of purely kinematic, gauge invariant objects, which in their most basic form have 
cyclically ordered  external legs and are called primitive amplitudes \cite{Bern:1994fz,DelDuca:1999rs}.
Cyclic ordering leads to a simplified dependence
on kinematic quantities, and as a result of this, primitive amplitudes are well suited for the application of on-shell techniques such as
BCFW recursion \cite{Britto:2005fq}, and the most developed formulations of unitarity techniques (such as those based on 
generalised unitarity \cite{Bern:1994zx,Bern:1997sc,Britto:2004nc,Forde:2007mi} and the numerical 
OPP reduction procedure \cite{Ossola:2006us,Ellis:2007br,Giele:2008ve,Berger:2008sj}).
Knowledge of how the colour information is reintroduced
can be used to calculate only those primitive amplitudes which will contribute to a particular order in a $1/N_c$ expansion, which
is particularly useful when computing one-loop high-multiplicity amplitudes since the very time consuming yet numerically 
small, sub-leading
colour parts can be neglected. Colour decompositions have also recently been applied to non-abelian gauge theories 
which are spontaneously broken \cite{Dai:2012jh}.
 Different methods of dealing with colour quantum 
numbers, such as Monte Carlo summation over fixed external colours \cite{Caravaglios:1998yr, Draggiotis:1998gr,Mangano:2001xp,Maltoni:2002mq}, and colour dressed recursive techniques \cite{Duhr:2006iq,Gleisberg:2008fv}, 
are highly efficient at tree-level; a recent study has developed this technique for use with 
virtual amplitudes \cite{Giele:2009ui}. 

In this paper I consider relations between primitive amplitudes composed of many massless antiquark-quark pairs at tree level. 
No all-$n$ formula is known to relate these primitives to the full amplitude, but  examples have been worked out
at tree and loop level (also with external gluons) by equating coefficients of Feynman diagrams \cite{Ellis:2011cr, Ita:2011ar} -- in \cite{Ita:2011ar}
a  colour decomposition in terms of one-loop primitives involving up to six quarks and one gluon was given (which is sufficiently complicated
 that it has to be provided in an attached text file to the paper), see also \cite{Badger:2012pg}. 
A motivation of this paper is the hope that a better understanding these objects and the relations they satisfy
could lead to an all-$n$ colour decomposition in terms of them.

As a way of introducing the multi-quark primitive amplitudes, 
consider first the trace-based colour decomposition for tree-level $n$ gluon scattering amplitudes in terms of  fundamental $SU(N_c)$
matrices, $\lambda^a$, \cite{Berends:1987cv,Mangano:1987xk},
\be
\mathcal{M}^{\text{tree}}_{\text{gluon}}(g_1,g_2,\ldots,g_n) =\sum_{\mathcal{P}(2\ldots n)} \text{tr}(\lambda^1\lambda^{2}\ldots\lambda^{n})\,\ma_{\text{gluon}}(1,2,\ldots,n)\,, 
\label{gludecomp}
\ee 
where the sum is over all permutations $\mathcal{P}$ of $2\ldots n$.
The labelling  of the primitive amplitudes $\ma_{\text{gluon}}$  attests to the fact that the only Feynman diagrams
which contribute to them, when drawn in a planar fashion, have a cyclic ordering of the external legs
which corresponds to the label. The $\ma_{\text{gluon}}$ inherit many properties from this colour
decomposition, and relations exist between them  (which I review in Sec.~\ref{sec2}). In particular, the Kleiss-Kuijf (KK)
relations \cite{Kleiss:1988ne} reduce the number of linearly independent primitive amplitudes from the $(n-1)!$ in 
Eq.~\ref{gludecomp} to $(n-2)!$. Bern-Carrasco-Johansson relations \cite{Bern:2008qj} further reduce
 the number of independent primitives to $(n-3)!$, but I will not discuss this colour-kinematic duality in this paper, and will use
 the term independent to implicitly mean independent over the field of real numbers. 
  Primitive amplitudes can be obtained directly using colour-ordered Feynman rules 
(see e.g. \cite{Dixon:1996wi}) that assign purely kinematic factors to vertices between particles which are
antisymmetric under exchange of two of the legs. 

The multi-quark primitive amplitudes  I will consider in this paper can be defined 
 by considering the scattering of $n$ external massless antiquarks plus quarks
  in the adjoint representation of $SU(N_c)$  -- then the amplitude has the same colour decomposition 
  as Eq.~\ref{gludecomp}, and I study relations between the corresponding primitive amplitudes $\ma_{\text{quark}}\equiv\ma$.
 They inherit all of the relations that the all-gluon amplitudes do from this colour decomposition,
 but they are evaluated with different colour ordered Feynman rules, and the presence of
  quark lines which require quark number and flavour conservation at their interaction
vertices introduces an interesting structure which gives rise to additional
 relations between the primitive amplitudes. 

\begin{figure}
\includegraphics[width=10cm]{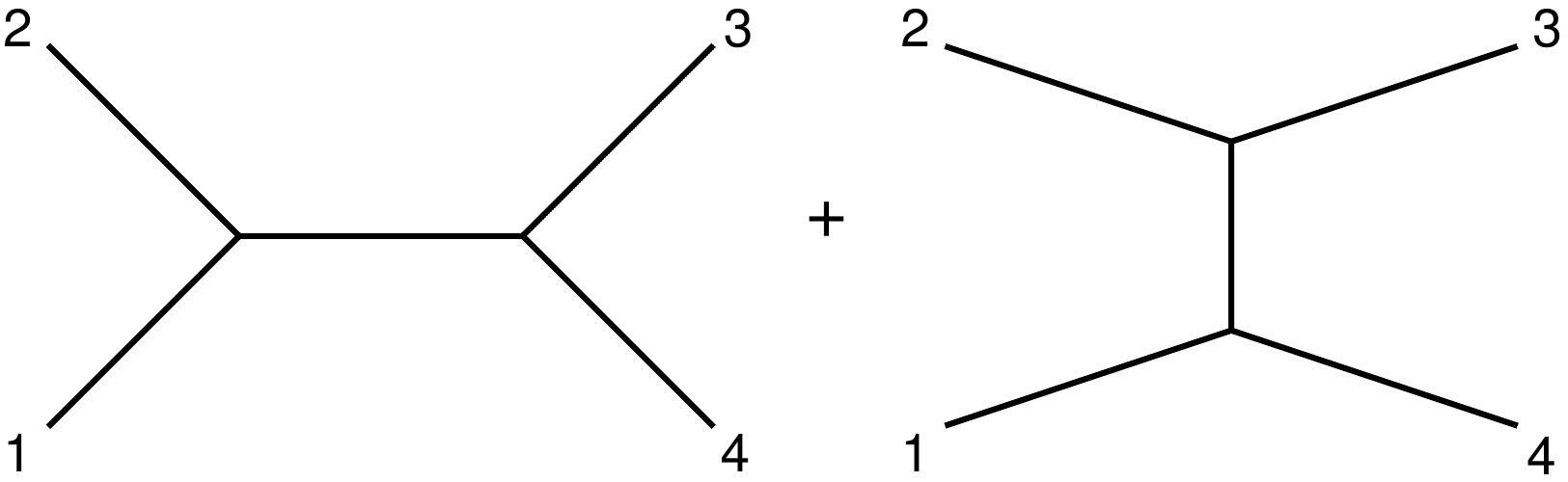}
\caption{The planar graphs contributing to a four particle primitive amplitude with cyclic ordering 1234.}
\label{cyclicplanes}
\end{figure}

As a simple example of this, consider the planar graphs which constitute a four particle primitive amplitude with the 
 cyclic ordering 1234 (see Fig.~\ref{cyclicplanes}). If all four particles are gluons, then both graphs
 are non-zero and contribute. When swapping gluons
 2 and 3 around to give the contributions to the primitive 1324, 
  the second diagram simply picks up a negative sign through the colour-ordered Feynman
 rules, whereas the kinematic structure of the 
 first diagram changes to something different. 
 Next consider the case where particle 1 is an up quark, particle 4 is an up antiquark, 
 particle 2 is a down quark, and particle 3 is a down antiquark. 
 The first graph is zero because it violates both flavour
and quark number at its vertices, whereas the second graph is allowed (the exchanged particle is a gluon); 
now when particles 2 and 3 are swapped around, the first
diagram is still zero (this time only due to flavour violation at the vertices) and the second diagram 
 picks up a negative sign. This
leads to a relation between the primitive amplitudes $1234=-1324$ which is not present in the all-gluon case. 
This type of relation was observed in \cite{Ita:2011ar} arising from non-trivial solutions to 
linear equations involving Feynman diagrams.
As a final example based on Fig.~\ref{cyclicplanes}, 
if particles 1 and 3 are an up quark and up antiquark, and particles 2 and 4 are a down quark
and down antiquark, then both diagrams are zero -- there is no planar way to connect the quarks of equal flavour.

In this paper, I will interpret the  relations described above as KK relations, and 
I show that in using them, an independent set of $(n-2)!/(n/2)!$  primitive amplitudes can be found
 for the case when all quarks have different flavours.
These amplitudes can be constructed using Dyck words (named after the German mathematician Walther von Dyck), which are strings of equal numbers of the letters $X$ and $Y$ such that the number of $X$s is greater
than or equal to the number of $Y$s in any initial segment of the string.

This paper is organised as follows. In Sec.~\ref{sec2} I will review the familiar 
properties of multi-quark primitive 
amplitudes and
introduce a useful graphical way of representing them.
In Sec.~\ref{sec3}  I will describe a correspondence between Dyck words and non-zero primitive amplitudes,
and provide a proof that the number of primitive amplitudes can be reduced to an independent set of size
$(n-2)!/(n/2)!$ using the KK relations. 
I conclude in Sec.~\ref{sec4}.

\section{Multi-quark primitive amplitudes}
\label{sec2}

Consider the $n$ particle tree-level scattering amplitude which has $n/2$ massless antiquark-quark pairs. I focus on the 
case where all the pairs have a distinct flavour (where  the term flavour can be used loosely in the
sense that two quark lines which have different helicities have different `flavour' even if they are both, up-type, down-type, etc.), and
will discuss at the end of this section the equal flavour case.
I  adopt the convention that odd momentum labels are assigned to antiquarks, even momentum labels are assigned
to quarks, and that the pairs of $\bar{q}q$ of equal flavour are $(1\to2), (3\to4), \ldots, (n-1\to n)$, where I use the notation $(\bar{q}\to q)$
to indicate a particular quark line. A colour decomposition when
these quarks are in the adjoint representation is,
\be
\mathcal{M}^{\text{tree}}_{\text{\bf adj}}(\overline{q}_1,q_2,\ldots,\overline{q}_{n-1},q_{n}) =\sum_{\mathcal{P}(2...n)} \text{tr}(\lambda^1\lambda^2\ldots\lambda^n)\,\mathcal{A}(1,2,\ldots,n-1,n)\,, \label{qadj}
\ee 
where as discussed in the introduction, the purely kinematic objects $\ma$ are planar, cyclically ordered, primitive amplitudes.
This decomposition is to be compared to the one when the (anti)quarks are in the fundamental 
representation with colour indices $(\bar{i})i$, \cite{Mangano:1988kk},
\be
\mathcal{M}^{\text{tree}}_{\text{\bf fund}}(\overline{q}_1,q_2,\ldots,\overline{q}_{n-1},q_{n}) =\sum_{\alpha\in S_{n/2}} \delta_{\bar{i}_1 \alpha_2}\delta_{\bar{i}_3 \alpha_4}\ldots\delta_{\bar{i}_{n-1} \alpha_n}\,\,\mb_{\alpha}\,, \label{qtree}
\ee 
 where the sum runs over all permutations $\alpha=(\alpha_2,\alpha_4,\ldots,\alpha_n)$ of quark indices $(i_2,i_4,\ldots,i_n)$. 
 The colour factors are strings of delta functions, and the $\mb_{\alpha}$ are
  purely kinematic functions (often called colour-ordered amplitudes, but they are not cyclically ordered).
 Factors of $1/N_c$ associated with a particular permutation $\alpha$ have been absorbed into the definition of the $\mathcal{B}_\alpha$.
As already mentioned, the $\mathcal{B}$ can be expressed in terms of the $\ma$ by solving linear systems of equations
defined by Feynman diagram expansions, 
but no all-$n$ formula is known.
 
Some of the relations between the $\ma$ in Eq.~\ref{qadj} are familiar from gluon amplitudes. 
They are gauge invariant, they are invariant under cyclic permutations of $1\ldots n$, they possess a reflection symmetry
$\ma(1,2,\ldots,n)=(-1)^n\ma(n,\ldots,2,1)$,
and they satisfy the KK relation, which may be written in the form,
\be
\ma(1,\{\beta\},2,\{\alpha\})= (-1)^{n_\beta}\sum_{\text{OP}\{\alpha\}\{\beta^T\}}\ma(1,2,\{\alpha\}\{\beta^T\}) \,,
\label{kk}
\ee
where $\{\alpha\}\cup\{\beta\}=\{3,\ldots,n-1,n\}$, $\{\beta^T\}$ is the set $\{\beta\}$ with the ordering of the elements reversed,
$n_\beta$ is the number of elements in $\{\beta\}$, and
 $\text{OP}\{\alpha\}\{\beta^T\}$ stands for `ordered permutations', which are the shuffle product of the elements of the sets $\{\alpha\}$ and $\{\beta^T\}$ -- 
 all permutations of the union of the two sets which keep fixed the ordering of the $\alpha_i$ within $\{\alpha\}$ and the
 $\beta_i$ within $\{\beta^T\}$.
\begin{figure}
\includegraphics[width=15cm]{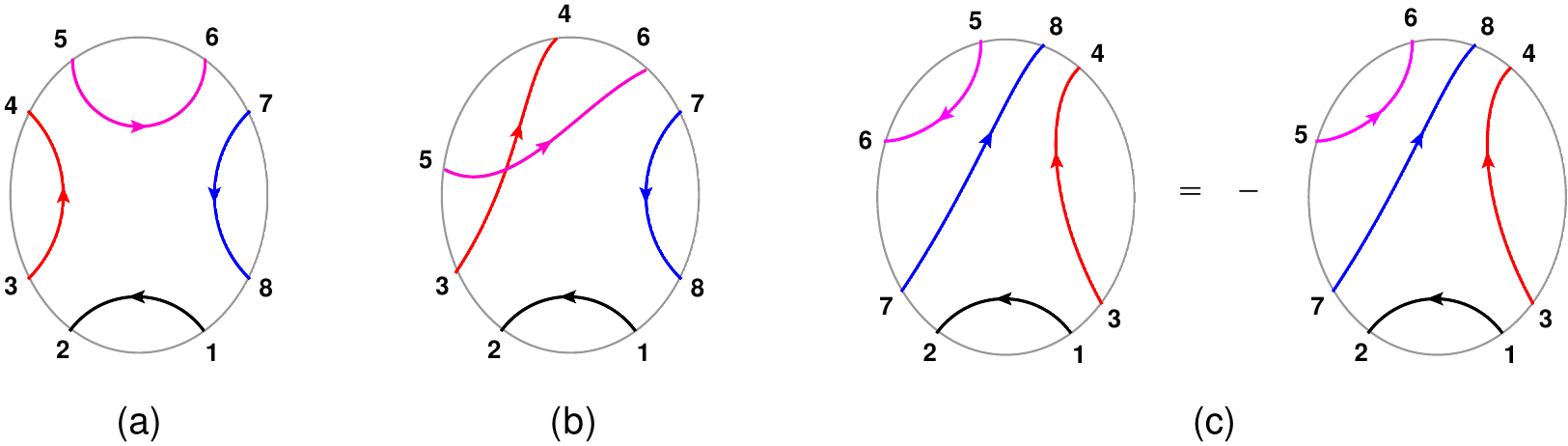}
\caption{(a)  Quark line graph for the permutation $\sigma=(12345678)$; (b) an example of a permutation
giving rise to crossed quark lines -- the corresponding primitive amplitude is zero; (c) a relation between primitive amplitudes.}
\label{firstdiags}
\end{figure}

Consider the set of primitive amplitudes with the labels for antiquark 1 and quark 2 fixed adjacent to each other in the order 12, 
but allowing for all permutations
of the labels $3\ldots n$ -- I call this set the $\as$ basis (I discuss the implications of fixing this choice of labels
in Sec.~\ref{sec4}). All other primitive amplitudes can be expressed in terms of elements of this set 
through Eq.~\ref{kk}.
A useful way to represent the $\ma$ graphically is shown in Fig.~\ref{firstdiags}.
A light grey circle is drawn to indicate the edge of the plane (for clarity, it does not mean any kind of trace is taken), 
and the quark labels are written clockwise around this circle in 
the order dictated by the particular permutation. Quark lines are then drawn to join $(1\to2)$, $(3\to4)$, etc..
These quark line graphs make clear the structure of the Feynman diagrams (computed using colour-ordered rules)
 which contribute to a given $\ma$ -- they are the tree diagrams which arise from joining the quark lines together with gluons in all
 distinct  planar ways.
They also make it easy to see that some $\as$ with certain permutations of $3\ldots n$ are zero. These are
the ones where quark lines cross, since there is no planar way in which to connect the crossed antiquark--quark pairs. 
An example of such a permutation is shown in Fig.~\ref{firstdiags} (b). 
\newline
Fig.~\ref{firstdiags} (c) shows a relation between primitive amplitudes in the $\as$ basis (and which does not hold for all-gluon amplitudes),
\be
\ma(1,2,7,6,5,8,4,3)=-\ma(1,2,7,5,6,8,4,3)\,. 
\label{akkrln}
\ee
This is really a KK relation with many of the primitive amplitudes entering it being zero, so they
are not written explicitly in Eq.~\ref{akkrln}.
Writing the relation graphically using quark line graphs, and explicitly showing the zero primitive amplitudes (which have crossed
quark lines), Eq.~\ref{akkrln} becomes
\be
\includegraphics[width=16cm]{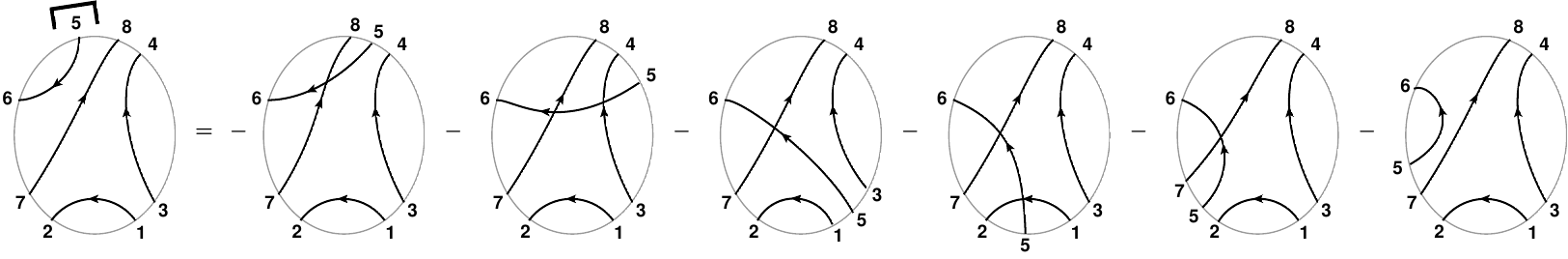}~~~~~
\label{kk1}
\ee
where
the bracket on the diagram on the LHS denotes which  leg which is to be associated with $\beta$ on the LHS of Eq.~\ref{kk}.
For this particular case, it is easy to understand the relation Eq.~\ref{akkrln} /\ref{kk1} in terms of the Feynman diagram expansion
of the two non-zero primitive amplitudes using the colour-ordered rules. This is because the quark line $(5\to6)$ is isolated
in a planar sense from the rest of the diagram, with the line $(7\to8)$ acting as a boundary 
to a zone in which it resides. In every contributing
Feynman diagram to $\ma(\ldots6,5\ldots)$, $(5\to6)$ connects only to $(7\to8)$, via a single gluon line, and the $6g5$ vertex 
can be flipped trivially, picking up a minus sign through the colour-ordered Feynman rules, 
to produce the corresponding Feynman diagram of $\ma(\ldots5,6\ldots)$.
This is to be compared with the case that arises from the KK relation 
\be
\includegraphics[width=16cm]{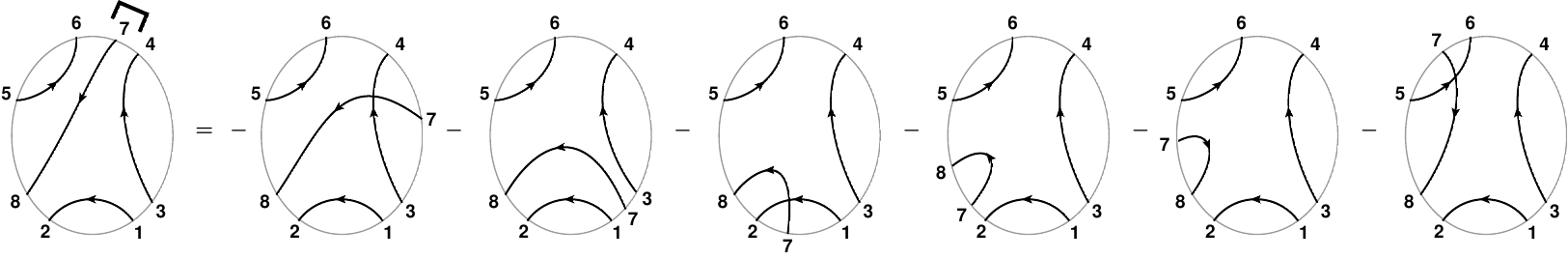}~~~~~
\label{kk2}
\ee
and relates four non-zero elements of the $\as$ basis (the ones without crossed quark lines). The interplay of the colour 
ordered Feynman diagrams in the expansions of each of these primitive amplitudes
is now more complicated, owing to the fact that the quark line $(7\to8)$ is not isolated and has to be joined
with gluons
to at least two of the other quark lines, and that this can be done in a number of different ways.
I will not consider Feynman diagram expansions of the primitive amplitudes in this paper, 
and will instead use the KK relations such as Eqs.~\ref{akkrln}/\ref{kk1},~\ref{kk2} directly in order to address the question of how
such relations impact on the number of independent multi-quark primitive amplitudes.

As presented above, the extra structure that having quark lines brings is to alter the KK relations so that
terms drop out, so that it is possible to re-express a primitive amplitude in the $\as$ basis in terms
of others which are all solely in the $\as$ basis. This never happens with all-gluon amplitudes, but having the additional
quark line structure introduces the possibility that the
amplitudes in a KK relation which are not in the $\as$ basis can be zero due to crossed quark lines (this is the case
with the $\ma(1,5,2,\ldots)$ in Eq.~\ref{kk1} and the $\ma(1,7,2,\ldots)$ in Eq.~\ref{kk2}). A question of interest
is how many independent $\as$ elements are left after all such relations are taken into account. An equivalent
way of posing this question is of asking which independent basis of $(n-2)!$ primitive amplitudes, chosen before quark line
considerations, is such that as many of the primitives as possible are zero once the quark lines are taken into account.
In starting with the $\as$ basis (chosen because it is initially a clearly independent set), the further relations between the elements of this basis
detail the extent to which this is not a basis of initial $(n-2)!$ independent primitives which maximises the number of zero
primitives when the conservation rules at quark-gluon vertices are enforced. It is this question (from the first point of view)
that I will address in Sec.~\ref{sec3}.
 
Up until this point, and  for the remainder of the paper (except for a brief discussion in Section~\ref{sec4}),
 I am considering the case where all quark lines are of different flavour. The amplitude where $n_e$ quark lines
have the same flavour, $\mathcal{M}^{\text{tree}}_{n_e}$, can be obtained using the all distinct flavour amplitude ($\mathcal{M}^{\text{tree}}_{n_e=0}$ in the below) by a permutation over quarks 
\be
\mathcal{M}^{\text{tree}}_{n_e}=\sum_{\mathcal{P}(q_1,q_2,\ldots q_{n_e})} (-1)^{\text{sgn}({\mathcal{P})}}\mathcal{M}^{\text{tree}}_{n_e=0}\,,
\label{flav}
\ee
where the sum is over all of the $n_e!$ permutations of  the equal  flavour quark indices, $q_1\ldots q_{n_e}$, and the $(-1)^{\text{sgn}({\mathcal{P})}}$ accounts for
Fermi statistics.
 An interesting observation is that when $n_e=n/2$, so that all quark lines are of equal flavour, none of the
primitive amplitudes in the $\as$ basis are zero. One way of seeing this is that when constructing
 a quark line graph, for any permutation of $3\ldots n$, it is always possible to find a way of joining up the odd numbers
 with even numbers without crossing lines (any join $(\text{odd}\to\text{even})$ is allowed as all flavours are the same).
 I prove this in App.~\ref{append0}.
 No primitive amplitudes
drop out of the KK relations, and so the number of independent
$\ma_{\,n_e=n/2}$ is $(n-2)!$, as it is for gluons. This fact, along with Eq.~\ref{flav} gives a bound on the number of independent
primitives  with $n_e=0$,
\be
\#\text{(independent } \ma_{\, n_e=0} \,)\ge (n-2)!/(n/2)!\,.
\label{numbind}
\ee
In the next Section, I will first discuss how to count all of
the possible non-zero graphs in the $\as$ basis, and then I will use KK relations (the additional quark line structure setting terms in them to zero)
to express this set in terms of one of it's subsets of size $(n-2)!/(n/2)!$. 
Eq.~\ref{numbind} then implies that this subset is independent.

\section{Dyck Words}
\label{sec3}

\begin{figure}
\includegraphics[width=16cm]{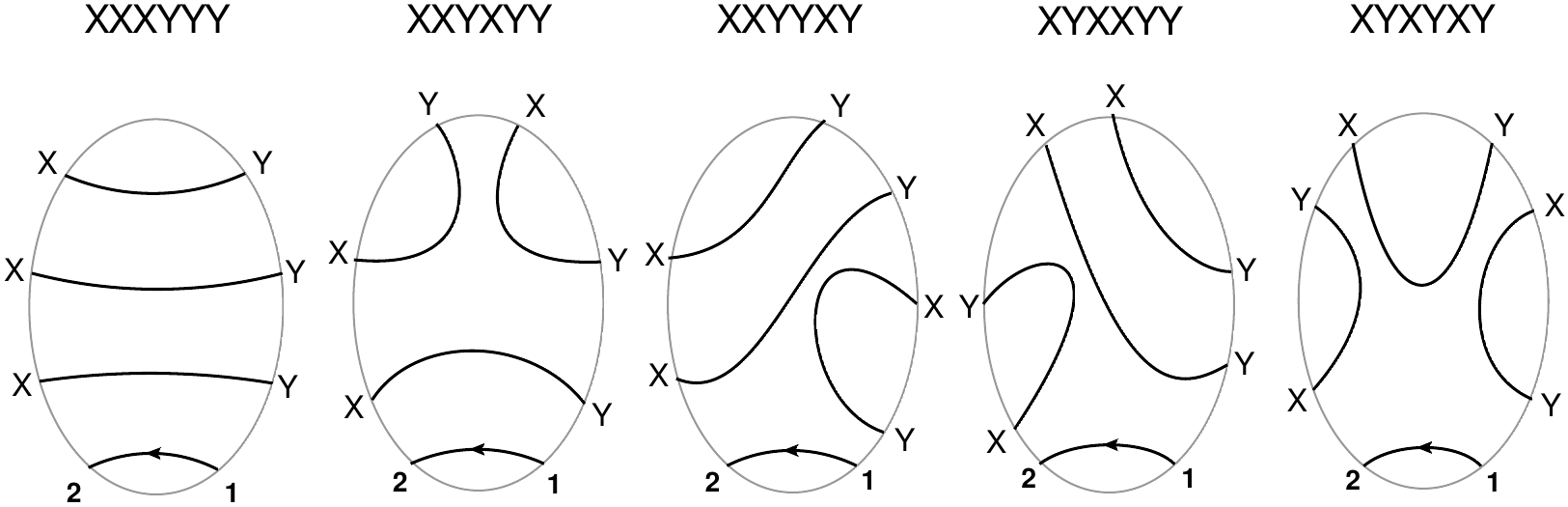}
\caption{Dyck words for $r=3$ (top row) and the quark line graph topologies they 
describe.}
\label{dycktree}
\end{figure}

A Dyck word is a string of length $2r$ consisting of  $r$ $X$s  and $r$ $Y$s,  such that the number of $X$s is greater
than or equal to the number of $Y$s in any initial segment of the string. The number of Dyck words of length $2r$
is given by the $r$th Catalan number, $C_r= (2r)!/(r+1)!r!$.
For example, for $r=3$ there are five Dyck words:
\be
XXXYYY~~XXYXYY~~XXYYXY~~XYXXYY~~XYXYXY\,.
\ee
A topology for a quark line graph can be associated with a Dyck word in the following way (see Fig.~\ref{dycktree}). First
draw in the line $(1\to2)$; then, moving clockwise from this line write  the Dyck word around the edge of the plane and each
time a $Y$ is encountered connect it with a line to the most recently written $X$ which has not already been connected.  
This procedure does not require any quark lines to cross, and
the Dyck words of length $2r$ provide all possible non-crossing topologies of  the $r=n/2 -1$  quark lines coming from the
$\sigma$ permutations in the $\as$ basis for $n$ quark scattering. That this is true can be seen from
the interpretation of Dyck words as
strings of correctly nested parentheses: $X\to$ `(' and $Y\to$ `)'. Identifying these parentheses 
with the two ends of a quark line it is clear that the correct nesting requirement is equivalent to avoiding crossed lines.

For each of these topologies, the quarks of different flavours can be assigned in $r!$ ways, and each quark line
can be directed in one of two ways, so the number of non-zero amplitudes in the $\as$ basis
is
\be
2^{r} \, r!\, C_{r} = 2^{\frac{n}{2}-1}(n-2)!/(n/2)! \,.
\ee
However, not all of these primitive amplitudes are independent. I shall now show that an independent subset can be chosen
defined as those primitives
 where the quark lines are all oriented in the same direction as the quark line $(1\to2)$. That is, when reading clockwise around
 the quark line graph, starting at antiquark 1 (or reading the label in $\as$ from left to right) the antiquark of each quark line
 is encountered before the quark.
 This removes the
 factor of $2^{\frac{n}{2}-1}$ in the above equation, and
  is a consequence of using KK relations to 
 expresses the other primitives in terms of members of this oriented subset (and other primitives of the form 
 $\ma(1\,\sigma'\,2\,\sigma)$ which are zero, where $\sigma\cup\sigma'=\{3,4,\ldots,n\}$ and $\sigma'\ne\emptyset\,$). 
As an example, the oriented primitives corresponding to the topologies shown in Fig.~\ref{dycktree} would have
antiquark labels assigned to the $X$s and quark labels assigned to the $Y$s.
In the following, I will use the term orient to mean re-express a primitive amplitude with wrongly directed quark lines 
in terms of primitive amplitudes where the
quark lines point in the same direction as the line $(1\to2)$.
I want to emphasise that this is not related to the charge parity equation,
\be
\ma(\ldots i^\pm_{\bar{q}}\ldots j^\mp_{q} \ldots )=-\ma(\ldots i^\pm_{q}\ldots j^\mp_{\bar{q}} \ldots)\,,
\ee
which assigns a different momentum and helicity $(\pm)$ to the quark $q$ on either side of the equation.

\begin{figure}
\includegraphics[width=8.5cm]{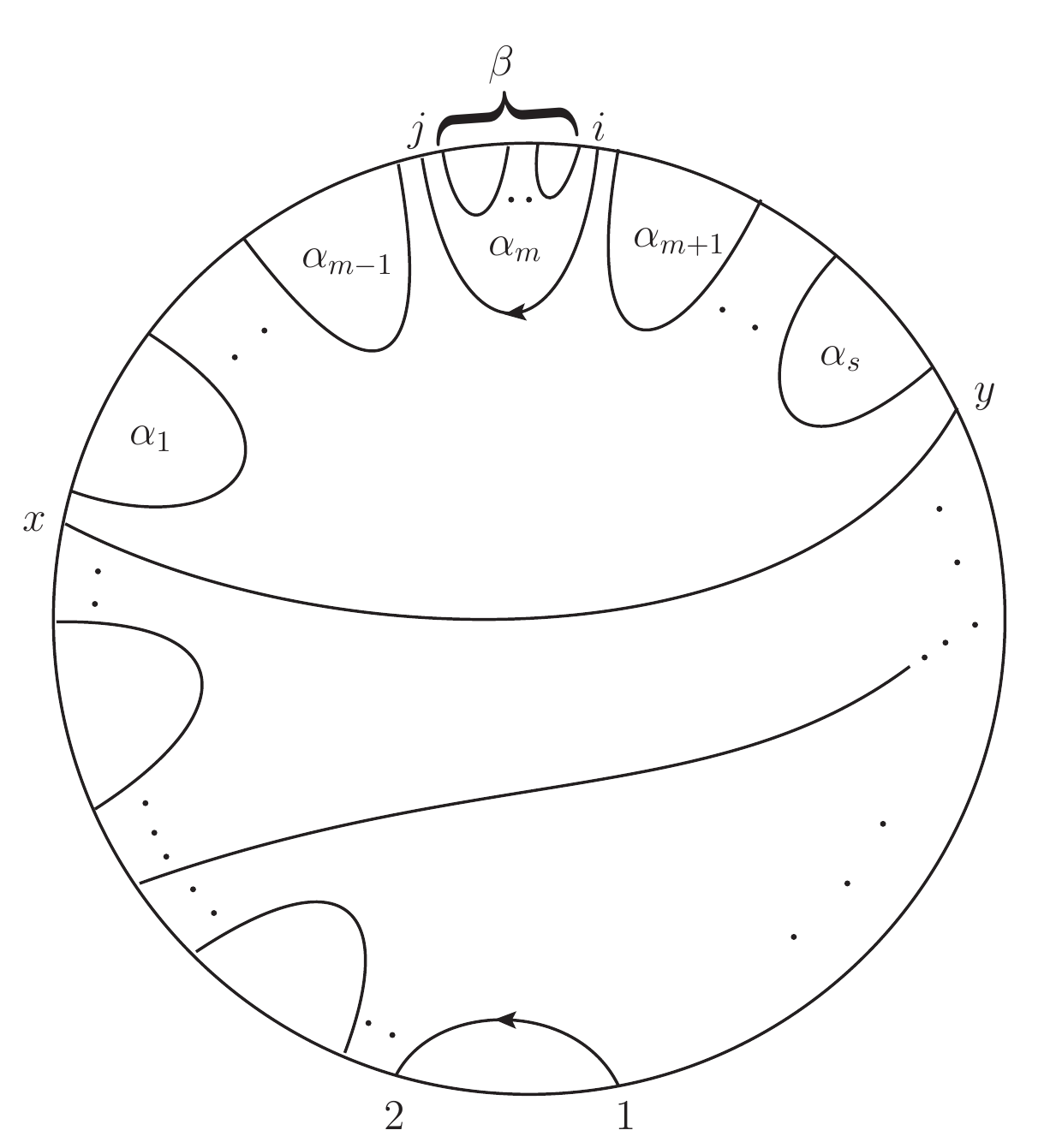}
\caption{A quark line graph for a primitive multi-quark amplitude, showing the general structure
of a  zone with a boundary of the quark line with ends labelled $x$ and $y$, and with sub-zones
 $\{\alpha_1\}\ldots\{\alpha_{s}\}$. The orientation of the boundary $(i\to j)$ of sub-zone $\{\alpha_m\}$ is explicitly
shown to be in the wrong direction, and the substructure of $\{\alpha_m\}$ is also explicitly shown and labelled $\{\beta\}$.}
\label{gendiag}
\end{figure}

Fig.~\ref{gendiag} depicts a zone bounded by the quark line with ends labelled $x$ and $y$ (direction not specified)
 within a generic primitive amplitude. (The ellipses outside
this zone stand for any number of quark lines, which may or may not straddle this zone -- a few example lines are shown, their direction is not
 specified). 
Within the zone
are $s$ sub-zones $\{\alpha_1\}\ldots\{\alpha_{s}\}$ which consist of a quark line boundary
and in general contain further substructure in the form of more quark lines (their boundaries are drawn in Fig.~\ref{gendiag}, but 
without specifying their
direction  for the sub-zones $\{\alpha_{1}\},\ldots,\{\alpha_{m-1}\},\{\alpha_{m+1}\},\ldots,\{\alpha_{s}\}$). The boundary of the sub-zone $\{\alpha_{m}\}$ 
is wrongly oriented, 
and this sub-zone is  shown explicitly broken down into its boundary (the line $(i\to j)$) and its substructure $\{\beta\}$.

I prove by induction that this zone can be oriented, by which I mean all of the quark lines inside the boundary
$(x-y)$ (and not the boundary itself) can be oriented.
For this, the following identity, a consequence of the KK relations and
 valid for multi-quark primitive amplitudes with different flavour quark lines (and which I prove in App.~\ref{append1}) is useful,
\be
\ma(\ldots x\,\{\alpha_1\}..\{\alpha_{m-1}\}\,j\,\{\beta\}\,i \,\{\alpha_{m+1}\}..\{\alpha_{s}\} \, y\ldots) = ~~~~~~~~~~~~~~~~~~~~~~~~~~~~~~~~~~~~~~~~~~~~~~~~~~~~~~~~~~~~~ \nonumber\\
-\sum_{c=1}^{m}\bigg[ \sum_{\text{OP}\{D_c\}\{E\}}\bigg(\sum_{\text{OP}\{A_c\}\{B\}}\ma(\ldots x\,\{\alpha_1\}..\{\alpha_{c-1}\}\,i\,\overbrace{\underbrace{\{\alpha_{c}\}..\{\alpha_{m-1}\}}_{\{A_c\}}\underbrace{\{\beta^T\}}_{\{B\}}\,j}^{\{D_c\}} \,\overbrace{\{\alpha_{m+1}\}..\{\alpha_{s}\}}^{\{E\}}  \,y\ldots) \bigg)\bigg]\,, \nonumber \\
\label{their}
\ee
where $\{\alpha_i\}$ are the sub-zones, and it should be remembered that some of the permutations induced by 
$\text{OP}\{A_c\}\{B\}$ and  $\text{OP}\{D_c\}\{E\}$ will be 
ones with crossed quark lines and will give rise to primitive amplitudes that are zero.

A zone which contains $k=1$ quark lines is oriented trivially 
as follows from  Eq.~\ref{their} with s=1, $\{\beta\}=\emptyset$,
\be
\ma(\ldots x \,j \,i \,y \ldots)= -\ma(\ldots x \,i\, j\, y \ldots)\,,
\ee
which is just the situation discussed beneath Eq.~\ref{kk1}, since the quark line $(i\to j)$ is completely isolated from the rest of the diagram
by the boundary $(x-y)$.
Now assume that a zone containing $k-1$ quark lines can be oriented. I will prove that
a zone containing $k$ quark lines can be orientated.  

If these $k$ quark lines are arranged so that the number of sub-zones $s=1$, the zone can be oriented as follows from Eq.~\ref{their} with $s=1$,
\be
\ma(\ldots x\,\,j\,\{\beta\}\,i \, y\ldots) = -\ma(\ldots x\,\,i\,\{\beta^T\}\,j \, y\ldots) \,.
\ee
This has correctly oriented the boundary of the sub-zone, which is all that is needed here (and in the following), for any substructure inside sub-zones
can be oriented by assumption by treating the boundary of the sub-zone as the boundary of a new zone containing $k'<k$ quark lines.
Now I assume that the case where the $k$ quark lines are arranged into $s-1$ sub-zones can be oriented, and
 show that the case with $s$ sub-zones can be oriented. To do this, apply Eq.~\ref{their}, in its general form,
\be
\ma(\ldots x\,\{\alpha_1\}..\{\alpha_{m-1}\}\,j\,\{\beta\}\,i \,\{\alpha_{m+1}\}..\{\alpha_{s}\} \, y\ldots) = &-& \ma(\ldots x\,\{\alpha_1\}..\{\alpha_{m-1}\}\,i\,\{\beta^T\}\,j \,\{\alpha_{m+1}\}..\{\alpha_{s}\} \, y\ldots) \nonumber \\
&+& \text{terms with smaller $s$}\,,
\label{solve}
\ee
which correctly orients the boundary of sub-zone $m$ up to other terms which by assumption can be oriented. Eq.~\ref{their} can be further applied
for any $m$, $1\le m \le s$, until all $s$ sub-zones have correctly oriented boundaries. This concludes the proof that the quark lines
contained within the full zone can all be oriented.

Finally, a special case of the above 
is the zone which contains $k=n/2-1$ quark lines and which has boundary $(1\to2)$ which is by definition
correctly oriented. Orienting this zone orients the full primitive amplitude. 

The reduced set of $(n-2)!/(n/2)!$ amplitudes obtained with the above method are
\be
\{\,\ma(1,2, \sigma) \,|\, \sigma \in \text{Dyck}_{\,n/2-1}[X_{\tau_1}...X_{\tau_{n/2-1}}Y\,...\,Y]\,\}\,,
\label{theamps}
\ee
where Dyck$_{\,r}$ means all Dyck words of length $2r$ composed of the $X$s and $Y$s, 
 with all $r!$ possible different labellings of the $X$s, i.e.
$\tau=(\tau_1,\ldots,\tau_{r})$ is a permutation of $(1,\ldots,r)$. 
The indices of the letter $Y$s  are determined by the  $X_{\tau_i}$ they are matched to via the method described 
at the beginning of this Section, which depends on the
particular Dyck permutation. If  a $Y$  gets
matched to $X_{\tau_i}$, then it is labelled as $Y_{\tau_i}$.
Finally the identification
$X_{1}\to3$, $Y_{1}\to4$, $X_{2}\to5$, $Y_{2}\to6$, $\ldots$ , $X_{n/2-1}\to n-1$, $Y_{n/2-1}\to n$ should be made. 

That this set of primitive amplitudes is independent follows from Eq.~\ref{numbind}.

\section{Discussion}
\label{sec4}

At the beginning of Sec.~\ref{sec2}, I made a choice of fixing the labels 1 and 2. This choice had an effect on the way in which
the paper progressed, since I have always been considering amplitudes in the $\as$ basis, and discussing relations between them.
Of course, now that an independent set is found in Eq.~\ref{theamps} -- call it $\{\ma(1,2,\text{Dyck})\}$ --
 (or this set where the role of $(1\to2)$ is switched with 
another quark line), all of the non-zero amplitudes which result from fixing, say, antiquarks 1 and 3 -- a $\ma(1,3,\sigma)$ basis -- can be
expressed in terms of this set. However, it is an interesting question as to which mixtures of 
primitive amplitudes from the $\{\ma(1,2,\text{Dyck})\}$, $\{\ma(3,4,\text{Dyck})\}$ etc.
bases are independent; addressing this question
 might help understand whether an all-$n$ colour decomposition in terms of primitive multi-quark 
amplitudes could be written down.

When carrying out the permutation sum in Eq.~\ref{flav} on $\{\ma(1,2,\text{Dyck})\}$ to obtain amplitudes where one or more quark lines have equal flavour,
the $n_e!$ primitive amplitudes at first sight also contain only quark lines which
 are correctly orientated, since Eq.~\ref{flav} only permutes the quark indices. 
 There is however an ambiguity since with equal flavour quark lines, the quark line graphs are not well defined -- 
 for instance if $(3\to 4)$ and
 $(7\to8)$ have the same flavour, then the graph for the primitive amplitude $\ma(12385674)$ could be drawn one of two ways,
 \be
 \includegraphics[width=7.5cm]{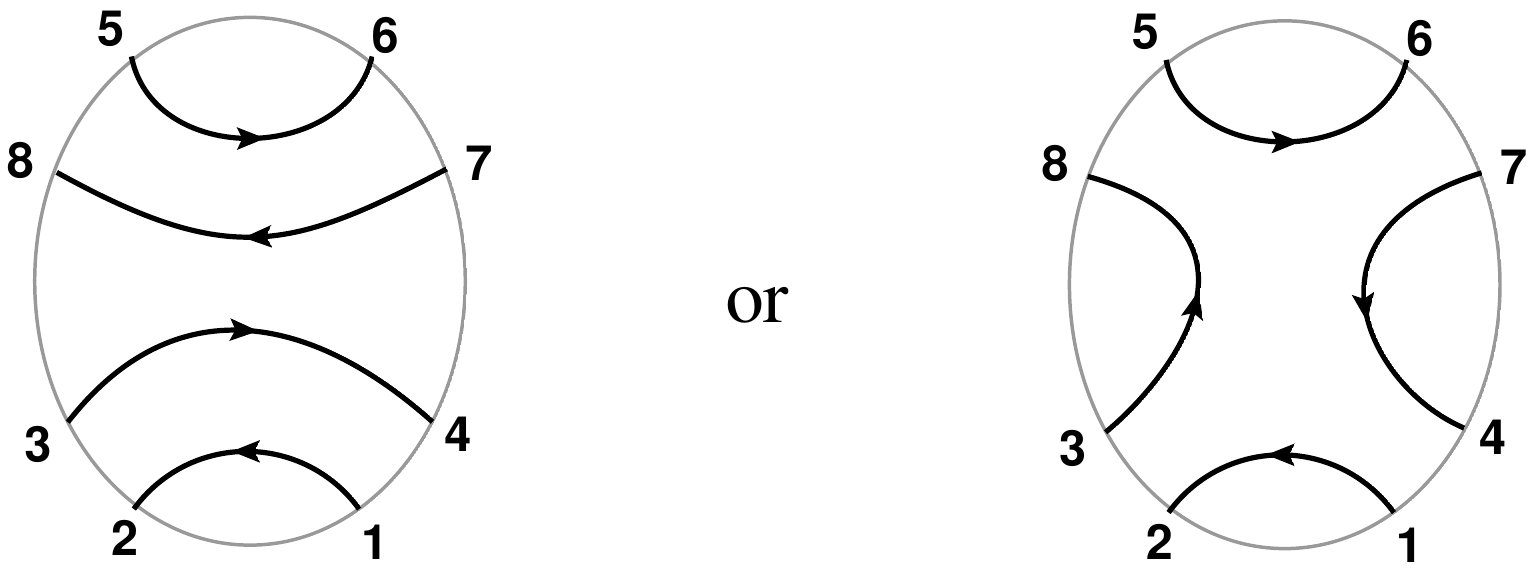}
 \ee
 The first of these diagrams is the one which would be drawn in the $n_e=0$ case, and
 it has the line $(7\to 8)$ incorrectly orientated.
 The second diagram has both quark lines $(3\to8)$ and $(7\to4)$ correctly oriented, and is what is obtained from the quark line graph for
 $\ma(12345678)$ by switching the labels $4\leftrightarrow8$ as in Eq.~\ref{flav}. The primitive $\ma(12385674)$ 
  receives contributions from Feynman diagrams 
 associated with both quark line graphs. It would be wrong to try and eliminate it  based on the first graph
 above. The KK relations are affected because fewer terms drop out of them -- when the quark lines
 $(3\to4)$ and $(7\to8)$ cross, which would  produces a zero primitive in the distinct flavour case.

 The addition of gluons to these pure-quark amplitudes is phenomenologically important -- contributions
 to collider processes with $n$ jets which involve quark lines are suppressed in colour by a factor of $1/N_c$ in comparison with
 the contribution, equal in $\alpha_s$, where the quark line is replaced by a gluon pair. The pure-quark primitive 
 amplitudes act as skeletons upon which to fix gluons, and the flavour structure, which this paper investigates,
 is not altered by their presence. I hope to think about mixed quark-gluon primitive amplitudes, in particular in the context of an
 all-$n$ colour decomposition, in the future.
 
Finally, it would be interesting (and useful in practical collider physics applications)
 to develop these ideas at one loop. Here, a relation similar to the KK relation at tree level
relates non-planar primitive amplitudes to planar ones. There is also a further division of multi-quark primitive amplitudes 
depending on whether quark lines turn left or right past the loop \cite{Bern:1994fz,  Ita:2011ar}. These amplitudes have already been
shown in \cite{ Ita:2011ar} to satisfy more relations than would all-gluon amplitudes, owing to their quark line structure. Such relations
 should be investigated in terms of the one-loop KK relations. Furthermore, the planar one-loop primitives retain the Dyck word structure
 that is seen here at tree level.
\newline

\noindent In conclusion, I have presented a proof that KK relations, along with flavour and quark number conservation,
reduce the number of independent purely multi-quark 
primitive amplitudes
 to $\nobreak{(n-2)!/(n/2)!}$ in the case when all quark lines have a different flavour. 
 I have given an explicit set of primitive amplitudes in Eq.~\ref{theamps} for any (even) value of $n$, which have a labelling of 
 external particles based on Dyck words.

\begin{acknowledgments}
I would like to thank Michelangelo Mangano and Franz Herzog for useful discussions.
This research was supported by ERC grant 291377.
\end{acknowledgments}

\appendix

\section{Non-crossing quark lines of equal flavour}
\label{append0}
In this Appendix, I prove the assertion in the text that when all quark lines are of equal flavour, a quark line graph
can always be drawn to connect quarks and antiquarks without any quark lines crossing. 
The proof is simple -- it relies on an argument which is similar to the construction of the Dyck topologies in Sec.~\ref{sec3}.
The quark line graph has some permutation of $n/2$ odd and $n/2$ even numbers written around the circle
 that represents the edge of the plane. Any odd number can be connected to any even number by a quark line, since
 the quarks and antiquarks all have equal flavour.
 Pick a starting point at an even number and move around the circle, 
 and each time an odd number is encountered, 
 connect it with a quark line to the most recently encountered even number that has not already been connected. 
This procedure leads to no crossed quarks lines,
proving the assertion.

\section{Proof of the identity Eq.~\ref{their}}
\label{append1}
I show how to apply KK relations in order to prove Eq.~\ref{their}, which is valid for purely multi-quark
scattering when all quark lines are of a different flavour. See also Fig.~\ref{gendiag}.
First use cyclic symmetry to write the LHS of Eq.~\ref{their} so that $x$ appears as the first label in the primitive
amplitude. Then apply the KK relation to $\{\alpha_1\}..\{\alpha_{m-1}\}$ (as indicated in the equation below),
\be
\ma(x\, \overbrace{\{\alpha_1\}..\{\alpha_{m-1}\}}^{KK}\,j\,\{\beta\}\,i \,\{\alpha_{m+1}\}..\{\alpha_{s}\} \, y\ldots) = \sum_{\text{OP}\{A\}\{B\}}\ma(\,x\,j \,\underbrace{ \{\alpha_{m-1}^T\}..\{\alpha_1^T\}}_{\{A\}}\,\,\underbrace{\{\beta\}\,i \,\{\alpha_{m+1}\}..\{\alpha_{s}\} \, y\ldots\,}_{\{B\}}\,  ) \,, \nonumber \\
\ee
and now use the fact that since the $\ma$ are multi-quark amplitudes  with quark lines of
 different flavour,  
these primitives are zero if an $\{\alpha_c\}$ straddles the $i$ (due to crossed quark lines). 
That is, the ordered permutations $\text{OP}\{A\}\{B\}$ can be split up 
 as follows,
\be
= \sum_{c=1}^{m} \bigg[ \sum_{\text{OP}\{C_c\}\{E\}} \bigg( \sum_{\text{OP}\{A_c\}\{B\}}   \ma(\,x\,j\, \underbrace{\{\alpha_{m-1}^T\}..\{\alpha_c^T\}}_{\{A_c\}} \, \underbrace{\{\beta\}}_{\{B\}} \,i\, \underbrace{ \{\alpha_{c-1}^T\}..\{\alpha_1^T\}}_{\{C_c\}} \,\underbrace{\{\alpha_{m+1}\}..\{\alpha_{s}\} \, y\ldots \,}_{\{E\}}\,) \,.   \bigg)\bigg] \nonumber \\
\label{a2}
\ee
Now consider a term with fixed $c$ and for one permutation of $\text{OP}\{A_c\}\{B\}$ in Eq.~\ref{a2}, define $\{\gamma_c\}=\{j\,\delta\}, \delta\in \text{OP}\{A_c\}\{B\}$, and then apply the KK relation to $\{\gamma_c\}$ in this term,
\be
\sum_{\text{OP}\{C_c\}\{E\}}  \ma(\,x\, \overbrace{\{\gamma_c\}}^{KK}\, \,i\, \underbrace{ \{\alpha_{c-1}^T\}..\{\alpha_1^T\}}_{\{C_c\}} \,\underbrace{\{\alpha_{m+1}\}..\{\alpha_{s}\} \, y\ldots \,}_{\{E\}}\,)  ~~~~~~~~~~~~~~~~~~~~~~~~~~~~~~~~~~~~~~~~~~~~~~~~\nonumber \\
= - \sum_{\text{OP}\{D_c\}\{F\}} \bigg( \sum_{\text{OP}\{C_c\}\{E\}}  \ma(\,x\, \,i\,  \overbrace{\{\gamma_c^T\}}^{\{D_c\}} \overbrace{\underbrace{ \{\alpha_{c-1}^T\}..\{\alpha_1^T\}}_{\{C_c\}} \,\underbrace{\{\alpha_{m+1}\}..\{\alpha_{s}\} \, y\ldots \,}_{\{E\}} }^{\{F\}}\,) \,\bigg)\, ~~~~~~~~~~~~~~~~  \nonumber  \\
= - \sum_{\text{OP}\{C_c\}\{F\}} \bigg( \sum_{\text{OP}\{D_c\}\{E\}}  \ma(\,x\, \,i\, \underbrace{ \{\alpha_{c-1}^T\}..\{\alpha_1^T\}}_{\{C_c\}} \underbrace{  \overbrace{\{\gamma_c^T\}}^{\{D_c\}}  \,\overbrace{\{\alpha_{m+1}\}..\{\alpha_{s}\} \, y\ldots \,}^{\{E\}} }_{\{F\}} \,)\, \bigg)\,  ~~~~~~~~~~~~~~~~  \nonumber  \\
= - \sum_{\text{OP}\{D_c\}\{E\}}  \ma(\,x\,\{\alpha_{1}\}..\{\alpha_{c-1}\}  \,i\,  \  \overbrace{\{\gamma_c^T\}}^{\{D_c\}}  \,\overbrace{\{\alpha_{m+1}\}..\{\alpha_{s}\} \, y\ldots \,}^{\{E\}}  \,) \,, ~~~~~~~~~~~~~~~~~~~~~~~~~~~~~~~~  
\label{a3}
\ee
where the third line is simply a rewriting of the permutation in the second line, and
where the last line follows through the KK relation with $\{\beta\}=\{\alpha_{1}\}..\{\alpha_{c-1}\}$. Substituting Eq.~\ref{a3} into Eq.~\ref{a2} yields the 
required identity, Eq.~\ref{their}.

\bibliography{planarqs.bib}

\end{document}